\documentclass{article}
\usepackage{style/spconf,amsmath,graphicx}
\usepackage{xcolor}
\usepackage{booktabs}
\usepackage{multirow}
\usepackage{caption}
\usepackage{subcaption}
\usepackage{enumitem}
\usepackage{cite}
\usepackage{amssymb}
\usepackage{pifont}
\newcommand{\cmark}{\ding{51}}%
\newcommand{\xmark}{\ding{55}}%

\title{Leveraging phone-level linguistic-acoustic similarity for utterance-level pronunciation scoring}
\name{Wei Liu$^{1,*}$\thanks{$^{*}$ This work was done during an internship at ByteDance.}  Kaiqi Fu$^{2}$, Xiaohai Tian$^{2}$, Shuju Shi$^{2}$, Wei Li$^{2}$, Zejun Ma$^{2}$ and Tan Lee$^{1}$}
\address{
$^{1}$Department of Electronic Engineering, The Chinese University of Hong Kong \\
$^{2}$ByteDance}

\begin{document}
\ninept
\maketitle
\begin{abstract}
Recent studies on pronunciation scoring have explored the effect of introducing phone embeddings as reference pronunciation, but mostly in an implicit manner, i.e., addition or concatenation of reference phone embedding and actual pronunciation of the target phone as the phone-level pronunciation quality representation. In this paper, we propose to use linguistic-acoustic similarity to explicitly measure the deviation of non-native production from its native reference for pronunciation assessment. Specifically, the  deviation is first estimated by the cosine similarity between reference phone embedding and corresponding acoustic embedding. Next, a phone-level Goodness of pronunciation (GOP) pre-training stage is introduced to guide this similarity-based learning for better initialization of the aforementioned two embeddings. Finally, a transformer-based hierarchical pronunciation scorer is used to map a sequence of phone embeddings, acoustic embeddings along with their similarity measures to predict the final utterance-level score. Experimental results on the non-native databases suggest that the proposed system significantly outperforms the baselines, where the acoustic and phone embeddings are simply added or concatenated. A further examination shows that the phone embeddings learned in the proposed approach are able to capture linguistic-acoustic attributes of native pronunciation as references.

\end{abstract}
\begin{keywords}
Linguistic-Acoustic Similarity, Phone Embedding, Goodness of Pronunciation, Pronunciation Scoring.
\end{keywords}

\section{Introduction}
\label{sec:intro}
Pronunciation scoring is an essential component of Computer Assisted Pronunciation Training (CAPT) \cite{witt2012automatic, fouz2015trends, chen2016computer, li2019improving,rogerson2021computer}. It is designed to automatically assess second language (L2) learners' speech pronunciations~\cite{neumeyer1996automatic, witt2000phone}. In general, the degree of proficiency/pronunciation level is measured as the amount of deviation of the L2 production from the reference native production. A typical assessment scenario is as follows: given a text prompt, the L2 learner is asked to read the text, and a scoring system is used to give a score based on the learner's speech production. 


Goodness of Pronunciation (GOP)~\cite{witt2000phone} was a commonly used feature in automatic pronunciation assessment, mispronunciation detection, and related tasks \cite{kanters2009goodness, strik2009comparing, huang2017transfer, ryu2017mispronunciation}. In a deep neural network (DNN) based system, GOP is computed as the ratio of log phone posterior probability between the canonical reference phone and the  hypothesized phone with the highest posterior probability~\cite{hu2015improved}. It gives a general-sense measurement of the pronunciation quality, i.e., a lower value of GOP indicates a higher possibility of mispronunciation. To improve the mispronunciation detection performance of GOP, various methods have been proposed. Transition probability between Hidden Markov Model (HMM) states was considered in  \cite{sudhakara2019improved} and a context-aware GOP score was investigated in~\cite{shi2020context}.


A comparison-based framework was investigated in~\cite{lee2012comparison, lee2013pronunciation, yue2017automatic}, where an utterance spoken by native speakers was adopted as a reference, and divergence-related features computed by dynamic time warping (DTW) between speech representations of native speakers and L2 learners were used to quantify the pronunciation deviation. However, parallel reference speech may not be available in real-world applications. Thus, it was considered to use phone embedding as the reference for pronunciation assessment \cite{lin2020automatic, lin2021deep, fu2022improving, gong2022transformer, chao20223m, zhang2021multilingual, lin2021neural}. One-hot representations of phoneme labels are fed into a trainable embedding layer to generate phone embedding vectors. The phone embeddings were used along with the corresponding phone-level acoustic embeddings for pronunciation score prediction. Addition \cite{lin2020automatic, lin2021deep, fu2022improving, gong2022transformer, chao20223m} and concatenation \cite{zhang2021multilingual, lin2021neural} of reference phone embedding and phone-level acoustic embedding are widely used methods to calculate phone-level pronunciation quality representation. The resultant representation is assumed to capture the deviation of non-native pronunciation from the reference production. However, either addition (add\_phone) or concatenation (concat\_phone) operation does not explicitly measure the degree of mismatch between what one native speaker pronounces (i.e., phone embedding) and how they actually pronounce (i.e., phone-level acoustic embedding). We hypothesize that explicit measurement of the degree of phone-level pronunciation deviation would better reflect L2 learners' pronunciation quality.

In~\cite{shao2022linguistic}, a linguistic-acoustic similarity based accent shift (LASAS) model was proposed for accent recognition. The accent shift is intended to capture the pronunciation variants of the same word in different accents. It is explicitly modeled by the similarity of acoustic embedding and aligned text anchor vectors. In the present study, we propose to use cosine similarity between a reference phone embedding and the corresponding acoustic embedding to explicitly measure the mismatch between standard and non-native pronunciation. A phone-level GOP pre-training process is developed to guide similarity-based learning for better initialization of the two embeddings. Lastly, a bottom-up hierarchical pronunciation scorer \cite{lin2021deep} is used to map a sequence of phone embeddings, acoustic embeddings along with the proposed similarity measures to predict the final utterance-level score. Experimental results show that the proposed system significantly improves the score prediction performance in terms of Pearson correlation coefficients (PCC) compared to its counterpart,  where phone and acoustic embeddings are simply added or concatenated. In addition, it is shown that the learned phone embedding can capture linguistic-acoustic characteristics of native pronunciation as references.

\begin{figure}[!t]
	\centering
	\includegraphics[width=\linewidth]{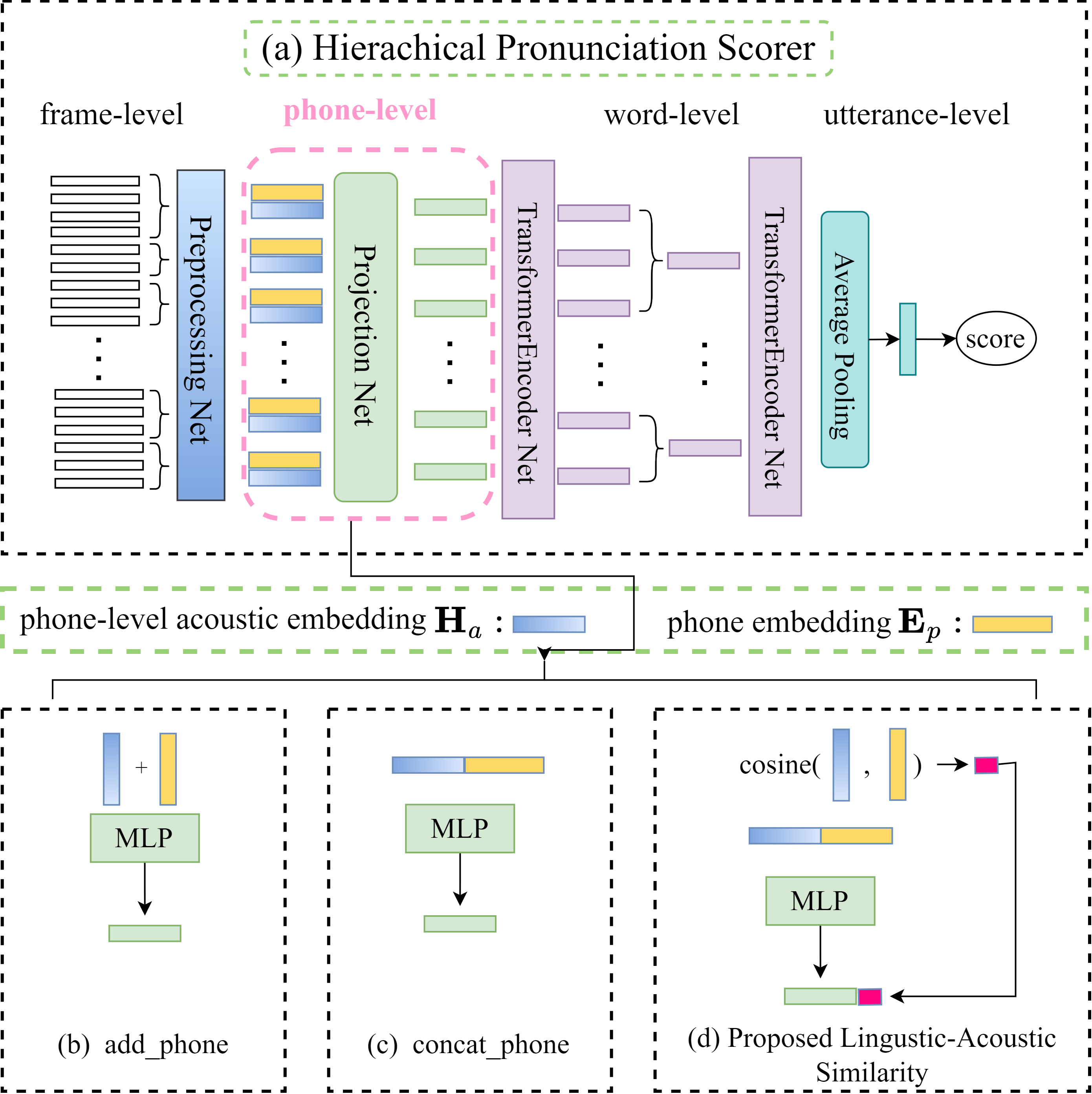}
	\caption{(a) illustrates the hierarchical architecture of the pronunciation scoring network. (b) and (c) show the two conventional methods of combining phone-level acoustic embeddings and phone embeddings. (d) shows the proposed linguistic-acoustic similarity method.}
	\label{fig:overview}
\end{figure}

\section{Method}
\label{sec:method}
In this section, we first give an overview of the hierarchical pronunciation scoring network, followed by an introduction to the proposed methods, including linguistic-acoustic similarity measure and phone-level GOP pre-training.

\subsection{Hierarchical Pronunciation Scorer}
\label{ssec:hierarchical}
As shown in Figure \ref{fig:overview} (a), 
the hierarchical pronunciation scoring network takes frame-level features as input, and aggregate and transform them into phone-level, word-level and utterance-level features layer by layer. The final output is a pronunciation score \cite{lin2021deep}.

Previous research \cite{lin2020automatic, lin2021deep, fu2022improving, gong2022transformer, chao20223m, zhang2021multilingual, lin2021neural} was focused on different implementations of representation learning for phone-level pronunciation quality aiming to model the deviation of L2 pronunciation from reference native pronunciation. Figures \ref{fig:overview} (b) and (c) depicts two such attempts. The vector in blue represents acoustic embeddings at phone-level after preprocessing network and the vector in yellow, i.e., the reference phone embedding, represents the native pronunciation of the current phone. 

Given a pair of read speech utterance and text prompt, an acoustic model is used to extract frame-level features (e.g., deep feature as in \cite{lin2021deep}) and phone-level alignment. 
Phone-level acoustic features are obtained by averaging the aligned feature frames of each phone segment, denoted as $\mathbf{X} \in \textbf{R}^{\text{D1} \times \text{N}}$. $\text{N}$ is the number of phones, and $\text{D1}$ is the feature dimension. 
Then, the phone-level acoustic feature and the corresponding phone ID are used as the input of the preprocessing network as shown in Eq. (1). A phone embedding layer encodes the phone ID $\mathbf{e}$ into phone embedding vectors, $\mathbf{E}_{p} \in \textbf{R}^{\text{D2} \times \text{N}}$. A fully connected (FC) layer projects the phone-level acoustic features $\mathbf{X}$ into the same dimension as phone embedding, denoted as $\mathbf{H}_{a} \in \textbf{R}^{\text{D2} \times \text{N}}$ . $\text{D2}$ denotes the embedding dimension.  $\mathbf{H}_{a}$ is termed as the phone-level acoustic embedding in this paper.
\begin{equation}
\label{eq:preprocessing}
\mathbf{H}_{a}, \mathbf{E}_{p} = \mathcal{F} (\mathbf{X}, \mathbf{e}),
\end{equation}
where $\mathcal{F}(\cdot)$ denotes the preprocessing network.

Subsequently, the phone-level projection network takes $\mathbf{H}_{a}$ and $\mathbf{E}_{p}$ as input and generates a pronunciation quality representation, denoted as $\mathbf{P}_{Q}$:
\begin{equation}
\label{eq:phone}
\mathbf{P}_{Q} = \mathcal{P} (\mathbf{H}_{a}, \mathbf{E}_{p}),
\end{equation}
where $\mathcal{P}(\cdot)$ represents the phone-level projection network. Specifically, it operates in two different ways:
\begin{equation}
\label{eq:add_phone_or_concat}  
\mathbf{P}_{Q} = \left\{
\begin{aligned}
\text{MLP}(\mathbf{H}_{a} + \mathbf{E}_{p}), ~~~~&\text{add\_phone}, \\
\text{MLP}([ \mathbf{H}_{a}; \mathbf{E}_{p} ]), ~~~~& \text{concat\_phone}.
\end{aligned}
\right.
\end{equation}
where $\text{MLP}$ refers to the multilayer perceptron function and $[;]$ denotes the concatenation of two vectors.



Finally, the utterance-level pronunciation score $\hat{y}$ is predicted as,
\begin{equation}
\label{eq:score}
\hat{y} = \mathcal{U} ( \mathcal{T}_{w}( \mathcal{A}_{w}( \mathcal{T}_{p}(\mathbf{P}_{Q}))) ) ,
\end{equation}

where $\mathcal{T}_{p}(\cdot)$ and $\mathcal{T}_{w}(\cdot)$ are the phone-level and word-level TransformerEncoder networks \cite{vaswani2017attention}, respectively. The $\mathcal{A}_{w}$ denotes the operation of averaging phone-level features to be word-level and $\mathcal{U}$ refers to the utterance-level output processing network.

\subsection{Proposed Method}
\subsubsection{Linguistic-Acoustic Similarity}
Eq.~(\ref{eq:add_phone_or_concat}) gives the two frequently used means of producing the phone-level pronunciation quality representation. Neither add\_phone nor concat\_phone explicitly measures the degree of mismatch between what the native speaker should pronounce (i.e., phone embedding) and how the L2 speaker actually pronounces (i.e., phone-level acoustic embedding).
To investigate the effect of modeling this pronunciation deviation in a more explicit manner, 
this study proposes a novel linguistic-acoustic similarity based learning method as illustrated in Figure \ref{fig:overview} (d). 
The phone-level pronunciation quality representation $\mathbf{P}_{Q}$ is calculated as in Eq.~(\ref{eq:phone}) but in a slightly different way:

\begin{equation}
    \label{eq:proposed_op}
    \mathbf{P}_{Q} = [\text{MLP}([ \mathbf{H}_{a}; \mathbf{E}_{p} ]); s ], 
\end{equation}
where $s$ denotes a linguistic-acoustic similarity measure, which is given by the cosine similarity between $\mathbf{H}_{a}$ and $\mathbf{E}_{p}$,

\begin{equation}
\label{eq:consine_similar}
s = cosine(\mathbf{H}_{a},\mathbf{E}_{p})
\end{equation}

The computation of utterance-level predicted score $\hat{y}$ remains unchanged.

\begin{figure}[!t]
	\centering
	\includegraphics[width=0.9\linewidth]{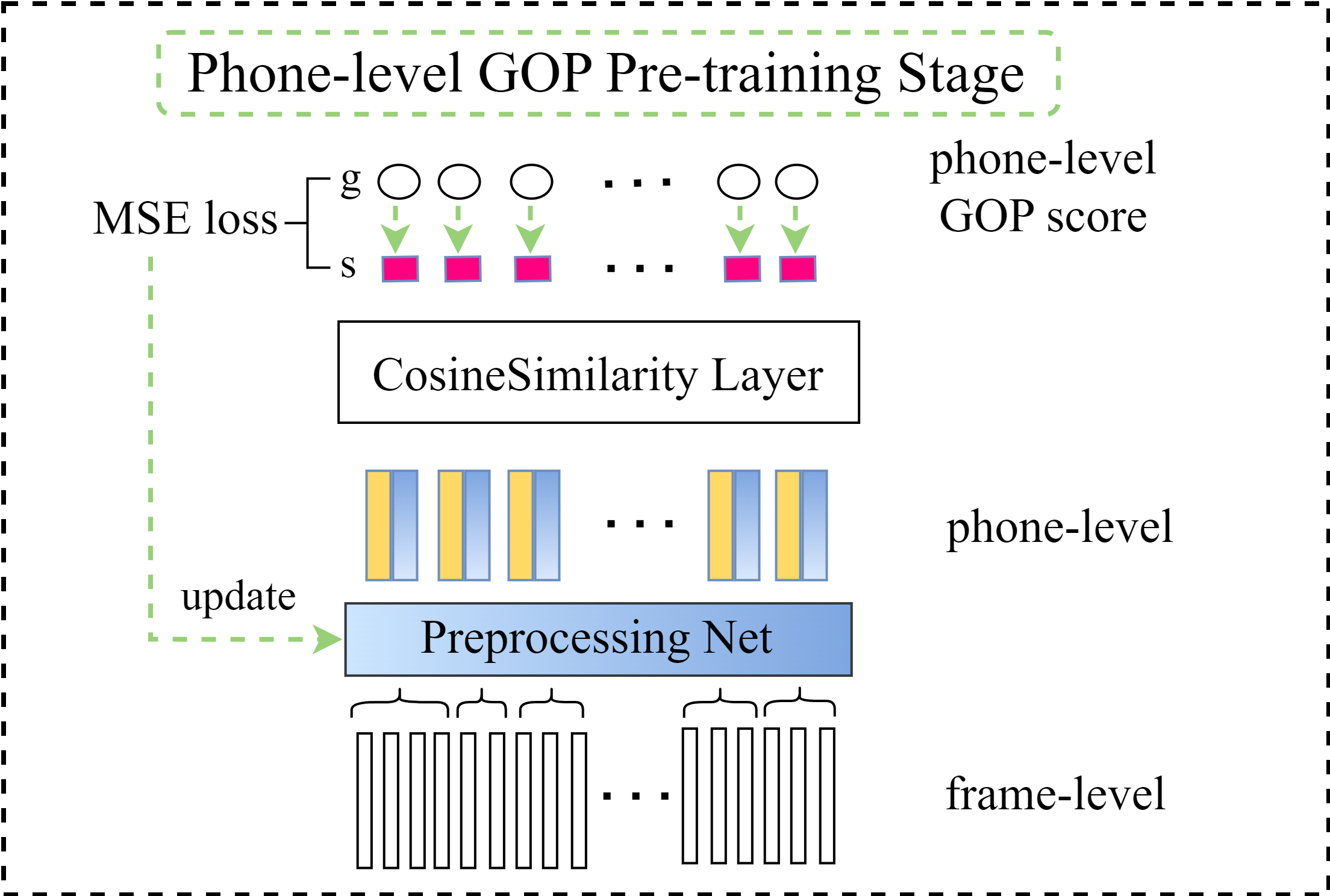}
	\caption{The diagram of the phone-level GOP pre-training stage. Note that only the preprocessing network is updated at this stage.}
	\label{fig:goppretrain}
\end{figure}

\subsubsection{GOP Pre-training}
\label{subsec:goppretrain}
GOP is widely used for measuring phone-level pronunciation quality of non-native speech, i.e., how close the pronunciation is to that of a native speaker \cite{witt2000phone}. To enable the phone-level linguistic-acoustic similarity to reflect pronunciation quality more accurately (e.g., a lower similarity indicates a higher possibility of mispronunciation), we use GOP score to guide the proposed similarity based learning.
Figure \ref{fig:goppretrain} shows the boxplot of the GOP pre-training of the phone-level preprocessing network.
Here, the used GOP score \cite{hu2015improved} for phone $p$ is calculated as follows:  
\begin{equation}
\label{eq:gop}
    GOP(p) = \mathcal{LPP}(p) - max_{q \in Q} \mathcal{LPP}(q),
\end{equation}
where $p$ is the phone in consideration and $Q$ is the whole phone set. $\mathcal{LPP}(p)$ is the log phone posterior and is computed as ${logp}(p|\mathbf{o}; t_s, t_e)$, where $t_{s}$ and $t_{e}$ are the start and end frame indexes of phone $p$, and $\mathbf{o}$ are the corresponding acoustic observations. Note that the phone-level time stamps and log posteriors are obtained using the extra acoustic model described in Section \ref{subsec:am}.

Mean squared error (MSE) between the cosine similarity $s$ and phone-level GOP score $g$ are used as the target loss function. We normalize both $s$ and $g$ into a range of [0,1]. Note that only the parameters of preprocessing network $\mathcal{F}(\cdot)$, as shown in Eq.~(\ref{eq:preprocessing}), are optimized at this stage, which results in an updated version of the phone-level acoustic embedding $\mathbf{H}_{a}$ and phone embedding $\mathbf{E}_{p}$.

\section{Experimental Setup}
\label{sec:exp_setup}
\subsection{Datasets}
Two L2 speech datasets are used in this study, namely Speechocean762 and ByteRate. Speechocean762 \cite{zhang2021speechocean762} is an open-sourced corpus designed for pronunciation assessment, in which 5,000 English utterances are collected from 250 learners. The corpus is split into train/test sets of equal size, each with 2,500 utterances from 125 English learners. Each utterance is rated by five experts in a range of 0 to 10, and the median value of the five scores is selected as the final score. ByteRate is an internal dataset at ByteDance, including a total of 10k utterances from 4k English learners. The train/dev/test sets are split as 3k/5k/2k, respectively. Each utterance is rated by three experts in a range of 0 to 4, and the final score is the average of the scores by all three experts. For both datasets, a higher rating indicates more native-like pronunciation and vice versa, and the scores are normalized into a range of 0 to 1. The first language (L1) of all L2 speakers is Mandarin.


\subsection{Model Configurations}
\subsubsection{Acoustic Model}
\label{subsec:am}
The deep feedforward sequential memory network and HMM, i.e., DFSMN-HMM, is adopted as the acoustic model~\cite{zhang2018deep}. 
DFSMN consists of 2 convolution layers and 24 FSMN layers followed by two FC layers. The input features are 39-dimension Mel-frequency cepstral coefficients (MFCCs). The acoustic model is trained on about 970-hour English speech, including an internal corpus of 10 hours of non-native English speech by L1 Mandarin learners and 960 hours of native English speech from the Librispeech corpus (Libri)~\cite{panayotov2015librispeech}. 512-dimensional deep feature is extracted from the penultimate layer of the acoustic model. 
The same acoustic model is used to force-align speech with the corresponding text prompt to obtain phone-level time stamps and compute GOP scores as shown in Eq. (\ref{eq:gop}). 



\begin{table}[t!]
\centering
\caption{The detailed structure of the proposed pronunciation scorer. LN refers to LayerNorm. Concat. is short for concatenation. Note that the time sequence information is omitted here.}
\begin{tabular}{c|cc|c}
\toprule
\textbf{Network}                                                                                     & \multicolumn{2}{c|}{\textbf{Structure}}                                                                                                  & \textbf{in $\times$ out size}            \\ \midrule \midrule
\multirow{2}{*}{Preprocessing}                                                            & \multicolumn{2}{c|}{{[}FC, LN, Tanh{]}}                                                                                         & 512 $\times$ 32                 \\ \cline{2-4} 
                                                                                          & \multicolumn{2}{c|}{{[}Embedding, LN, Tanh{]}}                                                                                  & 1 $\times$ 32                   \\ \hline
\multirow{4}{*}{Projection}                                                               & \multicolumn{2}{c|}{Concat. of $\mathbf{H}_{a}$ and $\mathbf{E}_{p}$}                                                                                & {[}32, 32{]} $\times$ 64        \\ \cline{2-4} 
                                                                                          & \multicolumn{1}{c|}{\multirow{2}{*}{MLP}}                                    & {[}FC, ReLU{]}                                   & 64 $\times$ 32                  \\ \cline{3-4} 
                                                                                          & \multicolumn{1}{c|}{}                                                        & FC                                               & 32 $\times$ 32                  \\ \cline{2-4} 
                                                                                          & \multicolumn{2}{c|}{Concat. with $s$}                                                                                           & {[}32, 1{]} $\times$ 33         \\ \hline
\multirow{3}{*}{\begin{tabular}[c]{@{}c@{}}Phone-level\\ TransformerEncoder\end{tabular}} & \multicolumn{2}{c|}{{[}LN, FC, Tanh{]}}                                                                                         & 33 $\times$ 32                  \\ \cline{2-4} 
                                                                                          & \multicolumn{2}{c|}{\multirow{2}{*}{\begin{tabular}[c]{@{}c@{}}att\_dim: 32, nhead: 4, \\ ff\_dim: 32, nlayer: 1\end{tabular}}} & \multirow{2}{*}{32 $\times$ 32} \\
                                                                                          & \multicolumn{2}{c|}{}                                                                                                           &                          \\ \hline
\multirow{3}{*}{\begin{tabular}[c]{@{}c@{}}Word-level\\ TransformerEncoder\end{tabular}}  & \multicolumn{2}{c|}{{[}FC, Tanh{]}}                                                                                             & 32 $\times$ 32                  \\ \cline{2-4} 
                                                                                          & \multicolumn{2}{c|}{\multirow{2}{*}{\begin{tabular}[c]{@{}c@{}}att\_dim: 32, nhead: 4, \\ ff\_dim: 32, nlayer: 1\end{tabular}}} & \multirow{2}{*}{32 $\times$ 32} \\
                                                                                          & \multicolumn{2}{c|}{}                                                                                                           &                          \\ \hline
Output                                                                                    & \multicolumn{2}{c|}{{[}FC, Sigmoid{]}}                                                                                          & 32 $\times$ 1                   \\ \bottomrule
\end{tabular}
\label{tab:config}
\vspace{-2mm}
\end{table}

\subsubsection{Pronunciation Scorer}
Table~\ref{tab:config} presents the detailed network configuration of the pronunciation scorer. 
The output dimension of preprocessing, projection, and TransformerEncoder networks is equal to 32 \cite{lin2021deep}.
For training the pronunciation scoring network, MSE between the predicted scores and the true scores is used as the loss function to be minimized. The training setups for  pronunciation scorer training and GOP pre-training (Section \ref{subsec:goppretrain}) are the same. The Adam optimizer is utilized with a learning rate of $0.002$ \cite{kingma2014adam}. The maximum number of epochs is set as 50, and early stopping is activated if the loss stops decreasing for seven consecutive epochs. It should be noted that the GOP pre-training stage does not involve any additional speech data. 




\section{Results and Analysis}
\label{sec:exp}
In this section, we present the experimental results of the proposed and baseline systems, analyze the effect of the GOP pre-training stage by comparing system performance using acoustic models trained on different amounts of non-native data, and further examine the linguistic-acoustic characteristics captured by the learned phone embeddings.
For performance evaluation, PCC between machine-predicted scores and human-predicted scores is calculated. 
\subsection{Performance of the Proposed and Baseline Systems}
\begin{table}[h]
\centering
\caption{The PCC results of the proposed and baseline systems. 
Note that the proposed and  baseline systems only differ in how to produce that phone-level pronunciation quality representation.
}
\resizebox{0.82\columnwidth}{!}{
\begin{tabular}{c|cccc}
\toprule
\multicolumn{1}{c|}{Datasets}     & \multicolumn{2}{c}{ByteRate} & \multicolumn{2}{c}{Speechocean762} \\ \midrule
\multicolumn{1}{c|}{GOP pretrain} & \xmark        & \cmark                 & \xmark           & \cmark                    \\ \midrule \midrule
add\_phone                        & 0.764   & 0.781            & 0.598     & 0.610               \\
concat\_phone                     & 0.763   & 0.823            & 0.618      & 0.652               \\ \midrule
\textbf{Proposed}                          & 0.825   & \textbf{0.858}   & 0.644      & \textbf{0.702}      \\ \bottomrule
\end{tabular}
}
\label{tab:overall}
\end{table}

\begin{figure*}[h]
  \centering
  
  \subfloat[
   add\_phone: vowels]
   {\includegraphics[width=0.25\textwidth]{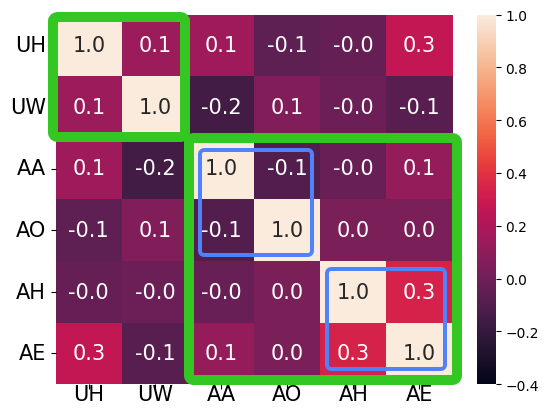}}
  \subfloat
    [Proposed: vowels]
   {\includegraphics[width=0.25\textwidth]{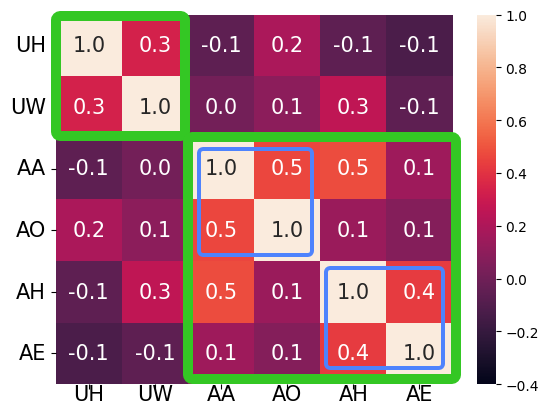}}
  \subfloat
    [add\_phone: consonants]
   {\includegraphics[width=0.25\textwidth]{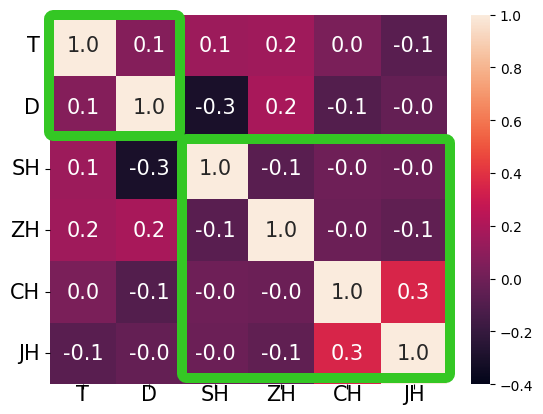}}
  \subfloat
    [Proposed: consonants]
   {\includegraphics[width=0.25\textwidth]{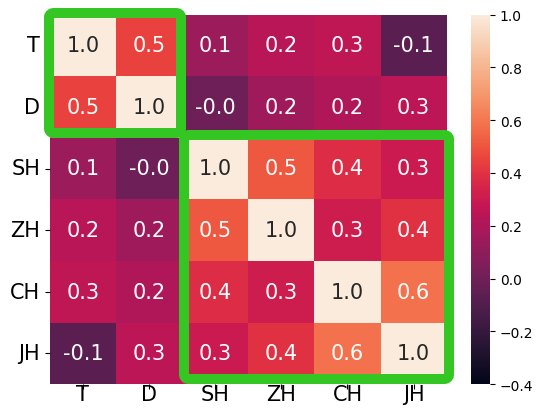}}
   
   
  
  \caption{Similarity heatmaps of phone embeddings between different phones (show-case examples). }
\label{fig:analyis}
\vspace{-1mm}
\end{figure*}


Table~\ref{tab:overall} presents the results of different systems on ByteRate and Speechocean762 datasets, respectively. We first examine the effectiveness of using phone-level linguistic-acoustic similarity for pronunciation assessment.
Compared to add\_phone and concat\_phone baselines, the proposed method improves the performance by a large margin on both ByteRate ($\uparrow$ 0.06 PCC) and Speechocean762 ($\uparrow$ 0.04 PCC) datasets.
The results suggest that the proposed linguistic-acoustic similarity can capture pronunciation deviation more effectively for pronunciation scoring.

We then examine the effectiveness of GOP pre-training for pronunciation assessment.
It is observed that, for all the three approaches, add\_phone, concat\_phone and the proposed system, the systems with GOP pre-training consistently outperform their counter-parts in terms of PCC. 
The proposed method achieves the best performance, with a PCC of 0.858 and 0.702 on ByteRate and Speechocean762 datasets, respectively.
Hence, we conclude that the proposed phone-level linguistic-acoustic similarity framework with GOP pre-training has a clear advantage over the baselines.

\begin{table}[h!]
\centering
\caption{The PCC results of the proposed system using two different acoustic models (AM) trained with different amount of L2 speech. }
\resizebox{0.88\columnwidth}{!}{
\begin{tabular}{c|cccc}
\toprule
Datasets          & \multicolumn{2}{c}{ByteRate} & \multicolumn{2}{c}{Speechocean762} \\ \midrule
GOP pretrain      & \xmark        & \cmark                & \xmark           & \cmark                    \\ \midrule \midrule
AM: Libri + 10h   & 0.825   & \textbf{0.858}   & 0.644      & \textbf{0.702}      \\
AM: Libri + 4000h & 0.860   & \textbf{0.893}   & 0.704      & \textbf{0.766}      \\ \bottomrule
\end{tabular}
}
\label{tab:am}
\vspace{-3mm}
\end{table}


\subsection{GOP Pre-training Stage: Less Can Be More }

Previous research has shown the benefits of introducing more non-native data in acoustic model training for L2 pronunciation assessment\cite{lin2021deep, gong2022transformer}. Acoustic models trained with both native and non-native data could provide more accurate phoneme segmentation of the L2 speech, hence better L2 phone representations for subsequent modeling processes. Unfortunately, non-native data of large size and high-quality annotation is not always available. In this study, we further examine how the proposed GOP pre-training process could help accommodate a lack of non-native speech data by conducting two more experiments which differ only in the amount of non-native data used during acoustic model training: 10 hours vs. 4,000 hours. The results are given in Table~\ref{tab:am}, which show that: (1) Including more non-native data in acoustic model training improves the system performance which is consistent with previous findings; (2) GOP pre-training is beneficial for both the 10 hour and 4,000 hours of non-native data conditions; and (3) The results of including 10h non-native data with GOP pre-training are comparable with results including 4,000 hour non-native data without GOP pre-training, suggesting that the GOP pre-training process could serve as an alternative when the amount of non-native data is limited.


\subsection{Linguistic-Acoustic Attributes of the Phone Embeddings}

In this section, we examine how (well) the learned phone embedding could relate to linguistic-acoustic attributes of its corresponding phoneme. In particular, we plot the similarity heatmaps between the phonemes based on the cosine similarity of their respective phone embeddings. The results for a group of vowels and a group of consonants are given in Figure~\ref{fig:analyis} as an example. In Figure~\ref{fig:analyis}, (a) and (c) show the results by the add\_phone approach, (b) and (d) by the proposed approach. Figure~\ref{fig:analyis} (b) clearly shows the pattern that the six vowels could be firstly divided into two clusters [UH, UW] and [AA, AO, AH, AE], and the second cluster could be further divided into two smaller clusters [AA, AO] and [AH, AE]. Specifically, the first two clusters, i.e., [UH, UW] and [AA, AO, AH, AE], differ in terms of vowel height, and the second in terms of tenseness. Similarly, in Figure~\ref{fig:analyis} (d), the six consonants seem to form two clusters [T, D] and [SH, ZH, CH, JH], with the sounds in the first cluster being plosives and those in the second one being fricatives or affricates. In either the vowel or the consonant group, similar patterns could not be observed from the phone embeddings obtained by the add\_phone approach. This shows that the phone embeddings learned by the proposed approach could reflect linguistic-acoustic attributes of their corresponding phonemes. Thus they are believed to provide more accurate reference representation of phone-level pronunciation.

\section{Conclusion}
\label{sec:conclusion}
In this paper, we proposed to use linguistic-acoustic similarity as additional feature to explicitly measure phone-level pronunciation deviation for pronunciation assessment. Moreover, a phone-level GOP pre-training stage was also proposed, which leads to better network initialization and more meaningful acoustic and phone embedding learning. The experiments conducted on both ByteRate and Speechocean762 datasets suggested that both linguistic-acoustic similarity and GOP pre-training
contribute to the performance improvement in terms of PCC. 
It is also shown that the phone embeddings learned in the proposed approach can capture linguistic-acoustic attributes of standard pronunciation as reference.
In the future, we plan to improve the system by using more contextualized acoustic features (e.g, wav2vec2.0) for the linguistic-acoustic similarity calculation. 

\bibliographystyle{style/IEEEbib}
\bibliography{refs}

\end{document}